\documentclass[twocolumn,showpacs,preprintnumbers,amsmath,amssymb]{revtex4}

\usepackage{graphicx}
\usepackage{dcolumn}
\usepackage{bm}
\usepackage{subfigure}

\makeatletter
\renewcommand{\@thesubfigure}{\Large\alph{subfigure}}
\renewcommand{\p@subfigure}{Fig.\space}
\renewcommand{\p@figure}{Fig.\space}
\makeatother %

\begin{document}

\preprint{APS/123-QED}

\title{Ion-induced secondary electron emission from K-Cs-Sb, Na-K-Sb and Cs-Sb \\ photocathodes and its relevance to the operation of gaseous avalanche \\ photomultipliers}

\author{A. Lyashenko}
\email{alexey.lyashenko@weizmann.ac.il}
\author{A. Breskin}%
\email{amos.breskin@weizmann.ac.il}
\author{R. Chechik}%
\affiliation{%
Dept. of Particle Physics, Weizmann Institute of Science; 76100 Rehovot, Israel}%
\author{T.H.V.T. Dias}
\affiliation{%
Physics Department, University of Coimbra, 3004-516 Coimbra,  Portugal}%

\date{\today}

\begin{abstract}
The operation of visible-sensitive gaseous- and, to some extent, vacuum-photomultipliers is critically affected by secondary electrons originating from ion impact on the photocathode. A simple method for indirect measurement of the effective ion-induced secondary-electron emission (IISEE) coefficient from the photocathode into a gas medium, $\gamma_+^{eff}$   was developed. The experimental results with visible-sensitive K-Cs-Sb, Na-K-Sb and Cs-Sb photocathodes, yielded $\gamma_+^{eff}$ - values between 0.02 and 0.03 in Ar/CH$_{4}$ (95/5) at 700 mbar; these are in good agreement with  theoretical calculations. The corresponding vacuum IISEE coefficients, $\gamma_+$, were estimated, based on a theoretical model, to be 0.47, 0.49 and 0.47 for K-Cs-Sb, Na-K-Sb and Cs-Sb photocathodes, respectively. The ratio of gas $\gamma_+^{eff}$ and vacuum $\gamma_+$ IISEE coefficients, calculated to be $\sim$0.06, is the fraction of secondary electrons surmounted the backscattering in the gas media.
\end{abstract}

\pacs{29.40.Gx, 29.40.Ka, 29.40.Cs, 85.60.Gz, 85.60.Ha}
\maketitle

\section{\label{sec:intro}Introduction}
The introduction of large-area position-sensitive photon detectors with single-photon sensitivity in the visible spectral range would lead to a significant progress in measuring light in numerous fields, such as particle-physics and astrophysics. At present, most commonly-used devices are vacuum photomultipliers (PMTs), with rather limited detection area and with bulky geometry due to mechanical constrains on the glass vacuum envelop. A possible reform could be the use of gas-filled photomultipliers (GPMs) operating at atmospheric pressure \cite{chechik:08}. They are expected to have large detection area, compact flat geometry, sub-mm 2D spatial resolution and fast (ns-scale) response. Efforts to realize this approach have been ongoing for almost 2 decades, gradually overcoming basic and technological obstacles related to the chemical and physical fragility of the photocathodes and their vulnerability to secondary processes \cite{edmends:88,peskov:99,balcerzyk:03}. One of the most critical impediments is the gas-avalanche Ion-Induced Secondary-Electron Emission (IISEE) process from the photocathode, which is the subject of the present work.

Visible sensitive GPMs comprise a photocathode (PC) coupled to a gaseous electron multiplier; the latter is preferably a cascade of hole-multipliers \cite{chechik:08}, e.g. Gas Electron Multiplier (GEM \cite{sauli:97}) and Microhole and Strips Plate (MHSP \cite{veloso:00}) elements; such cascaded multipliers are opaque to gas-avalanche-photons, thus preventing secondary photoemission from the photocathode.  The yet unsolved and most difficult issue, preventing high-gain operation of such devices, is the flow of avalanche ions from the amplification region to the PC [moerman:tes]. Impinging on the PC surface, these ions have a non negligible probability of releasing secondary electrons, which in turn initiate secondary avalanches known as ion feedback; the latter limits the detector's gain by diverging into discharges \cite{moermann:thes}.

The secondary electron emission into gas differs from that in vacuum as the electrons emitted from the PC are subject to back-scattering from gas molecules; the value of ion-induced secondary-electron emission (IISEE) coefficient in the gas media is lower than that in vacuum. The effect depends on the gas type, due to difference in the scattering cross-sections for various gases. The backscattering effect is smaller in gases with a complex molecular structure; e.g. it is very strong in atomic gases but is rather weak in organic compound gases.

The same backscattering effect is responsible for affecting the quantum efficiency (photoelectron emission) in the gas to be smaller than that in vacuum.  It should be noted, however, that because photoelectrons and ion-induced secondary electrons do not have the same energy, the corresponding backscattering effect may differ in size.  In GPMs, it is essential to maintain the two contradicting conditions: to allow for the highest possible quantum efficiency, while at the same time to reduce to minimum the ion feedback probability. For that purpose it is desirable to decrease the back-flow of avalanche ions to a level which, together with the given IISEE probability, will not cause gain divergence to discharge.  Extensive studies were carried out on ion back-flow reduction in cascaded electron multipliers; they have been reported elsewhere \cite{bondar:03, sauli:05, lyashenko:06, lyashenko:07}. These studies showed that, with an appropriate choice of the cascade elements and their operation conditions, the ion backflow (IBF) fraction can be as small as $3\cdot10^{-4}$, namely the number of ions back-flowing to the photocathode could be reduced down to about 30 for a total number of $10^5$ avalanche ions \cite{lyashenko:07}; this constitutes a record in ion blocking in gaseous detectors. On the other hand, the IISEE probability from bi-alkali PCs has not been reported yet. Its value is important for estimating the maximum attainable multiplication factors or, alternatively the   IBF fraction required for stable operation of gaseous GPMs.

In this work we investigated the IISEE from K-Cs-Sb, Na-K-Sb and Cs-Sb PCs, both experimentally and theoretically. Measurements were carried out with PCs coupled to a double gaseous electron multiplier (double-GEM). The ion-induced secondary emission probabilities were deduced from the experimental gain- curves' shapes of the multiplier. The experimental data were validated by a theoretical model for ion-induced secondary electron emission from solids.

\section{\label{sec:intro}Experimental setup and methods}
\begin{figure}
    \includegraphics[width=6cm]{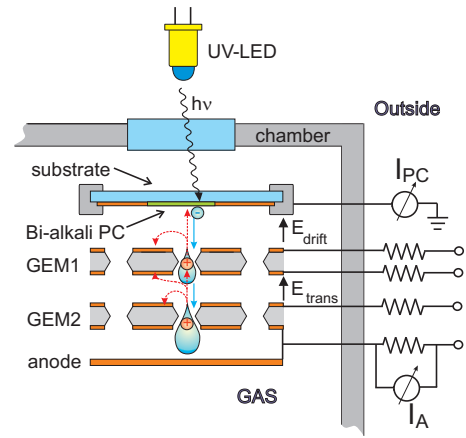}
    \caption{\label{fig:IISEE:setup} Schematic view of a double-GEM GPM with a semitransparent photocathode. Photoelectrons are extracted from the photocathode into the gas, they are focused into the holes of GEM1, multiplied and transferred into GEM2 holes for further multiplication. The avalanche ions (their possible paths are depicted by dotted arrows), in turn, drift back following the same electric field lines. The majority of ions are captured by the GEM electrodes and only a fraction reaches the PC. The GPM's gain and ion back-flow are established by recording currents at the anode (interconnected with the GEM2 bottom) and at the PC.}
\end{figure}

The GPM assembled for the IISEE studies comprised a double-GEM cascaded multiplier coupled to different visible-sensitive semitransparent PCs; the latter, bi- or mono-alkali ones, were vacuum deposited in a dedicated UHV system \cite{shefer:98, balcerzyk:03, moermann:thes} on a glass substrate (\ref{fig:IISEE:setup}). The GEM electrodes, of 28x28 mm$^2$ effective area, were produced at the CERN printed circuit workshop, from 50 $\mu$m thick Kapton foil with 5 $\mu$m Au-coated copper cladding on both faces; the holes were double-conical of 70/50 $\mu$m (outer/inner) diameter, arranged in a hexagonal pattern with a pitch of 140 $\mu$m. All the components of the multiplier were UHV-compatible, including the GEM electrodes; the latter are known to be compatible with bi-alkali PCs \cite{balcerzyk:03}. The GEM electrodes were mounted on 1 mm spaced ceramic frames.

The assembled multiplier was mounted in a UHV system, dedicated to preparation and studies of visible-sensitive PCs in vacuum and in gas media; the system comprises three chambers, in which PCs are prepared and characterized under different conditions. The system and the methodology of PC preparation and characterization with gas electron multipliers is described in \cite{shefer:98, moermann:thes, balcerzyk:03}. There is an option of sealing assembled GPMs \cite{moermann:thes}, which was not used in this study. The system maintains a base vacuum of $\sim10^{-10}$  mbar.

Following the introduction of the multiplier into the detector's chamber, the entire system was baked at 150$^0$C for 5 days in high vacuum. The PC was deposited and characterized under vacuum in the preparation chamber; next, both the preparation and detector's chambers were filled, through a purifier, with Ar/CH$_{4}$ (95/5) gas mixture to a pressure of 700 mbar. The PC was then transferred in-situ and placed in the detector's chamber, 8 mm above the multiplier, to constitute a GPM.

The GPM was irradiated continuously with a UV-LED (NSHU590A, Nishia Corp., 375nm peak wavelength) focused onto the PC by means of a small lens through a quartz window. The detector and the electric scheme are shown in \ref{fig:IISEE:setup}. Photoelectrons were transported into GEM1 holes under a drift field $E_{drift}$; following a gas-multiplication process in the holes, avalanche electrons were transferred under a transfer field, $E_{tans}$, into the second multiplier GEM2, for further multiplication; charges induced by the two-stage multiplication were collected on an anode, interconnected with the "bottom" face of GEM2 to assure full charge collection. The GEM electrodes were biased independently with HV power supplies of type CAEN N471A; the PC was kept at ground potential. The current after multiplication was recorded on the biased bottom GEM2/anode electrode as a voltage-drop on a 40 M$\Omega$ resistor, with a Fluke 175 voltmeter of 10 M$\Omega$ internal impedance (\ref{fig:IISEE:setup}). The combined resistance was 8 M$\Omega$, from which the anode current was calculated. The avalanche-induced currents were always kept well below 100 nA by attenuating the UV-LED photon flux, to avoid charging-up effects. The currents on grounded electrodes were recorded with a Keithley 485 picoamperemeter. The multiplication (gain) curves of the GPM were deduced from the ratio of the anode current ($I_{A}$) to the photocurrent emitted, without multiplication, from the PC ($I_{PC0}$). The avalanche-induced IBF fraction is defined as a ratio of the anode current $I_{A}$ to the avalanche ion-induced current at the PC $I_{PC}$ (see \ref{fig:IISEE:setup}).

\section{\label{sec:IISEE:effects}Theoretical evaluation of the ion-induced secondary emission effects}
\subsection{\label{sec:IISEE:theory}Estimation of the secondary emission coefficient}

In ionized gas mixtures, an effective process of charge exchange takes place, substituting ions of high ionization potentials, as they drift towards the PC, by ions with lower ionization potentials. As mentioned in \cite{sauli:77}, it takes between 100 and 1000 collisions for an ion to transfer its charge to a molecule having a lower ionization potential. Since the mean-free-path $\lambda$ for ion collisions with gas molecules is of the order of 10$^{-5}$ cm at room temperature and atmospheric pressure \cite{sauli:77}, one can assume that after a drift length of 10$^{-3} \cdot p^{-1}$ to 10$^{-2} \cdot p^{-1}$ cm, (where $p$ is the fraction of molecules with the lowest ionization potential in the mixture), the charge-exchange mechanism will leave only one species of ions drifting in the gas. In Ar/CH$_{4}$ (95/5) mixture used in this work, the distances for complete charge exchange are between 0.2 and 2 mm; they are therefore several times smaller then the 8 mm drift gap kept between the PC and the multiplier. The effective charge exchange in Ar/CH$_{4}$ was also confirmed experimentally in \cite{moermann:thes}, through the similarity of ion-induced secondary emission coefficients measured from K-Cs-Sb PC into CH$_{4}$ and Ar/CH$_{4}$. Hereafter in the calculations we will consider only CH$_{4}$ ions. In typical operating conditions of GPMs the electric field at the PC surface ($E_{drift}$) is 0.2-1 kV/cm \cite{breskin:02}. Under these conditions, and due to collisions in the gas,  the back-flowing ions have rather low kinetic energy (below 1 eV), which is too small for kinetic induction of secondary-electron emission \cite{kaminsky:65, raether:64}.

\begin{figure}
    \includegraphics[width=5cm]{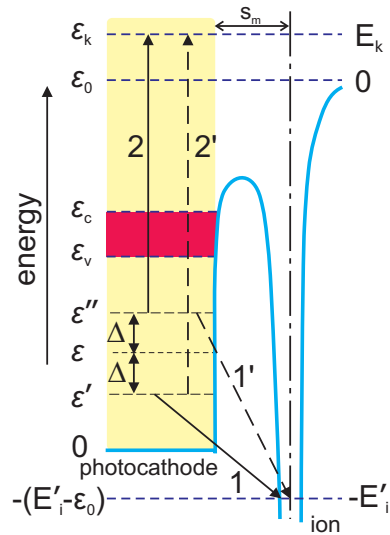}
    \caption{Energy-level diagram of the Auger neutralization process. Two sets of transitions are shown (1+2 and 1'+2'), to delineate the energy range in which the process is possible. The energies indicated on the left side of the drawing are inside the solid (e.g. bi-alkali PC) with zero at the bottom of the valence band; those on the right side of the drawing are outside the solid with zero level corresponding to vacuum level or to the energy of free electron at rest at an infinite distance from both ion and solid.}
    \label{fig:auger:diagram}
\end{figure}

The most favorable ion-induced electron emission process is the Auger neutralization process, as discussed in \cite{hagstrum:61}. The theory of Auger neutralization of noble gas ions at semiconductor surfaces was thoroughly described in \cite{hagstrum:61, takeishi:62} Therefore, we shall focus here only on the main aspects of this phenomenon; all notations used below are those of \cite{hagstrum:61}.
\begin{table*}
\begin{ruledtabular}
    \begin{tabular} {ll}
    Notation & Meaning \\  \hline
    $E_{i}^{'}$ & effective ionization energy at a distance $s_m$ from the PC surface\\
    $\varepsilon^{'}$, $\varepsilon^{''}$ & initial energies in valence band of electrons of electrons participating in Auger neutralization\\
    $\varepsilon_{v}$ & valence band maximum energy \\
    $\varepsilon_{c}$ & conduction band minimum energy \\
    $\varepsilon_{0}$ & vacuum level energy \\
    $\varepsilon_{k}$ & energy of the excited Auger electron inside the PC \\
    $E_k=\varepsilon_{k}-\varepsilon_{0}$ & kinetic energy of the excited Auger electron outside the PC \\
    \end{tabular}
    \caption{Notations used in the text and in \ref{fig:auger:diagram}}
    \label{tab:IISEE:notations}
\end{ruledtabular}
\end{table*}
In the vicinity of the PC surface, an ion induces polarization of the PC material, which can be formulated as an image charge. Due to the interaction of the ion with the image charge, the ionization potential of the ion shifts by
\begin{equation}
\label{eq:ion:screen}
      \Delta E_{i}=-\frac{(\kappa-1)\cdot e^2}{(\kappa+1)\cdot 4 \cdot s}
\end{equation}
where $\kappa$ is the dielectric constant of the PC material, $s$ is the distance between the ion and PC surface, $e$ is the electron charge. The resulting effective ionization potential is given by $E_{i}^{'}=E_{i}-\Delta E_{i}$ , where $E_{i}$ is the free space ionization potential (e.g. 12.6 eV for methane). As an ion with effective ionization potential $E_{i}^{'}$ approaches the surface of the PC, the probability to get neutralized by an electron from the valence band of the PC increases up to a maximum at a distance $s_m$ from the PC surface \cite{hagstrum:61}. The distance $s_m$ can be approximated as the average of the nearest-neighbor distance, $a_{nn}$, in the semiconductor (PC) and the molecular diameter of the gas (e.g. methane) molecule, $d_{gas}$:
\begin{equation}
\label{eq:sm}
      s_{m}=\frac{a_{nn}+d_{gas}}{2}
\end{equation}
The values of the parameters  $\kappa$,  $a_{nn}$, $s_{m}$ for some PC materials used in our calculations are listed in Table \ref{tab:IISEE:parameters}; the molecular diameter of methane is assumed to be 3.8 ${\AA}$ \cite{mohr:99}. The energy diagram of the electron transitions in the Auger neutralization process is depicted in \ref{fig:auger:diagram}, with the notations listed in Table \ref{tab:IISEE:notations}. The energies inside the solid are indicated on the left side of the diagram; those outside the solid are shown on its right side. The valence band extends from zero to $\varepsilon_{v}$; the conduction band minimum is at $\varepsilon_{c}$; the bands are separated by a gap of $\varepsilon_{c}-\varepsilon_{v}$; the vacuum level is at $\varepsilon_{0}$. Two electrons in the valence band, with initial energies $\varepsilon^{'}$ and $\varepsilon^{''}$, are involved in the Auger transition: one electron will neutralize the ion and occupy the vacant ground level of the ion; the other electron will be excited by the released energy and will jump to an energy state $\varepsilon_{k}$ in the conduction band. If it surmounts the surface barrier $\varepsilon_{0}$, it becomes an ion-induced secondary electron with an energy $E_{k}=\varepsilon_k-\varepsilon_0$. \ref{fig:auger:diagram} depicts two possible processes of this type, 1+2 and 1'+2'.

Energy conservation requires:
\begin{equation}
\label{eq:auger:conservation}
      \varepsilon^{'}+\varepsilon^{''}=2 \cdot \varepsilon=\varepsilon_k+\varepsilon_0-E_k=E_k+2 \cdot \varepsilon_0-E_i^{'}
\end{equation}
and by definition $\varepsilon_{k}=E_k+\varepsilon_{0}$. The maximal $(E_k)_{max}$ and minimal $(E_k)_{min}$ kinetic energy of the excited electron may now be evaluated from equation \ref{eq:auger:conservation}. The maximal kinetic energy is reached when $\varepsilon=\varepsilon_v$; it is given by:
\begin{align}
 \label{eq:auger:emax}
     (\varepsilon_{k})_{max}=E_i^{'}-\varepsilon_{0}+2 \cdot \varepsilon_{v}, \quad or \notag \\ \quad (E_k)_{max}=E_i^{'}-2 \cdot (\varepsilon_{0}-\varepsilon_{v})
 \end{align}
The minimum of the kinetic energy is reached when $\varepsilon=0$; it is given by:
\begin{eqnarray}
 \label{eq:auger:emin}
    (\varepsilon_{k})_{min}=E_i^{'}-\varepsilon_{0} \quad for \quad E_i^{'}-\varepsilon_{0}>\varepsilon_{c}; \nonumber\\  (\varepsilon_{k})_{min}=\varepsilon_{c} \quad for \quad E_i^{'}-\varepsilon_{0} \leq \varepsilon_{c}, \quad and \nonumber\\
    (E_k)_{min}=E_i^{'}-2 \cdot \varepsilon_{0} \quad for \quad E_i^{'}-\varepsilon_{0}>\varepsilon_{c}; \nonumber\\ (E_k)_{min}=0 \quad for \quad E_i^{'}-\varepsilon_{0} \leq \varepsilon_{c}
\end{eqnarray}

\begin{figure}%
\subfiguretopcaptrue
\subfigure 
{
    \label{fig:DOS:CsSb}
    \includegraphics[width=7cm]{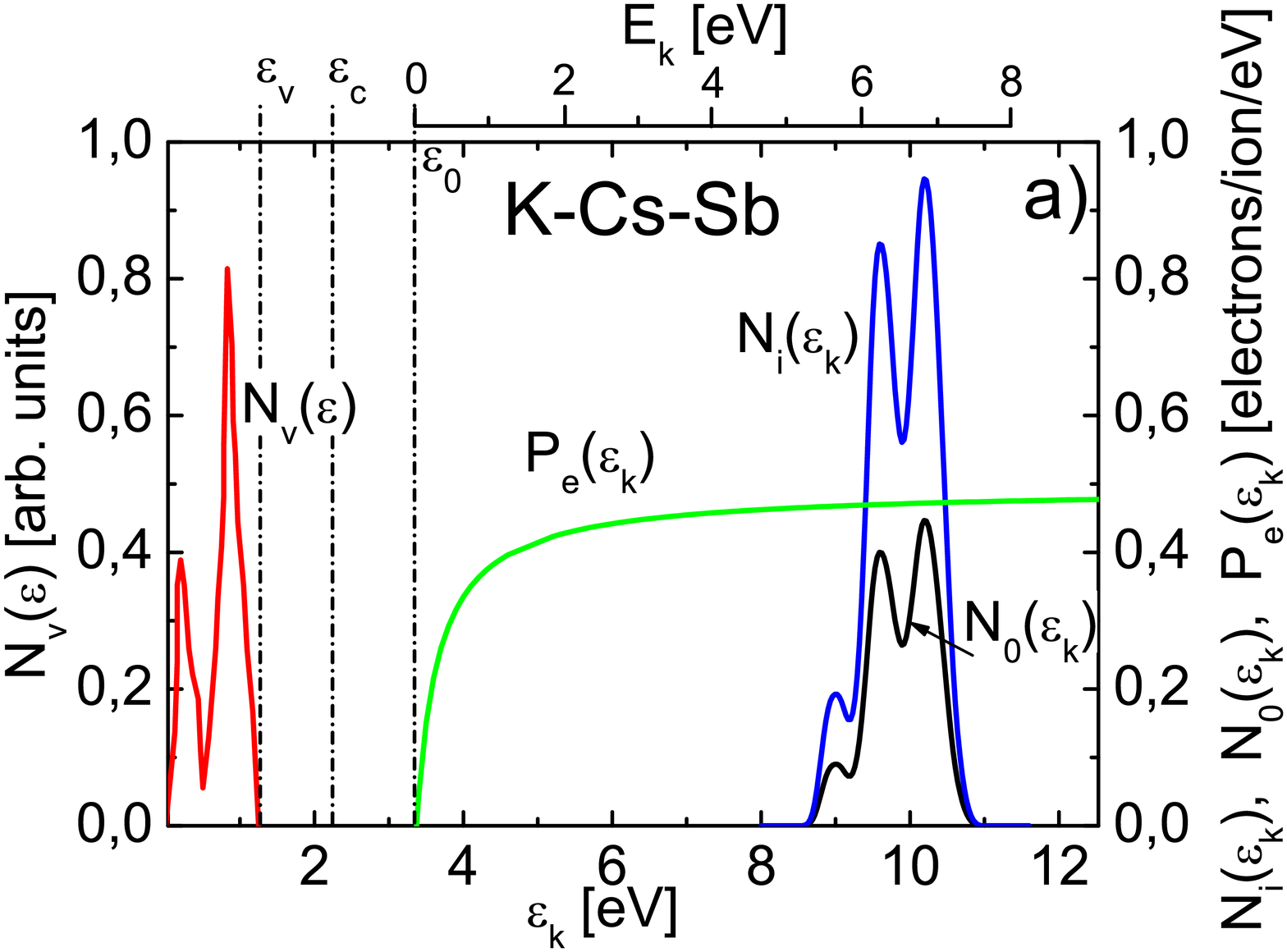}
} \hspace{1cm}
\subfigure 
{
    \label{fig:DOS:KCsSb}
    \includegraphics[width=7cm]{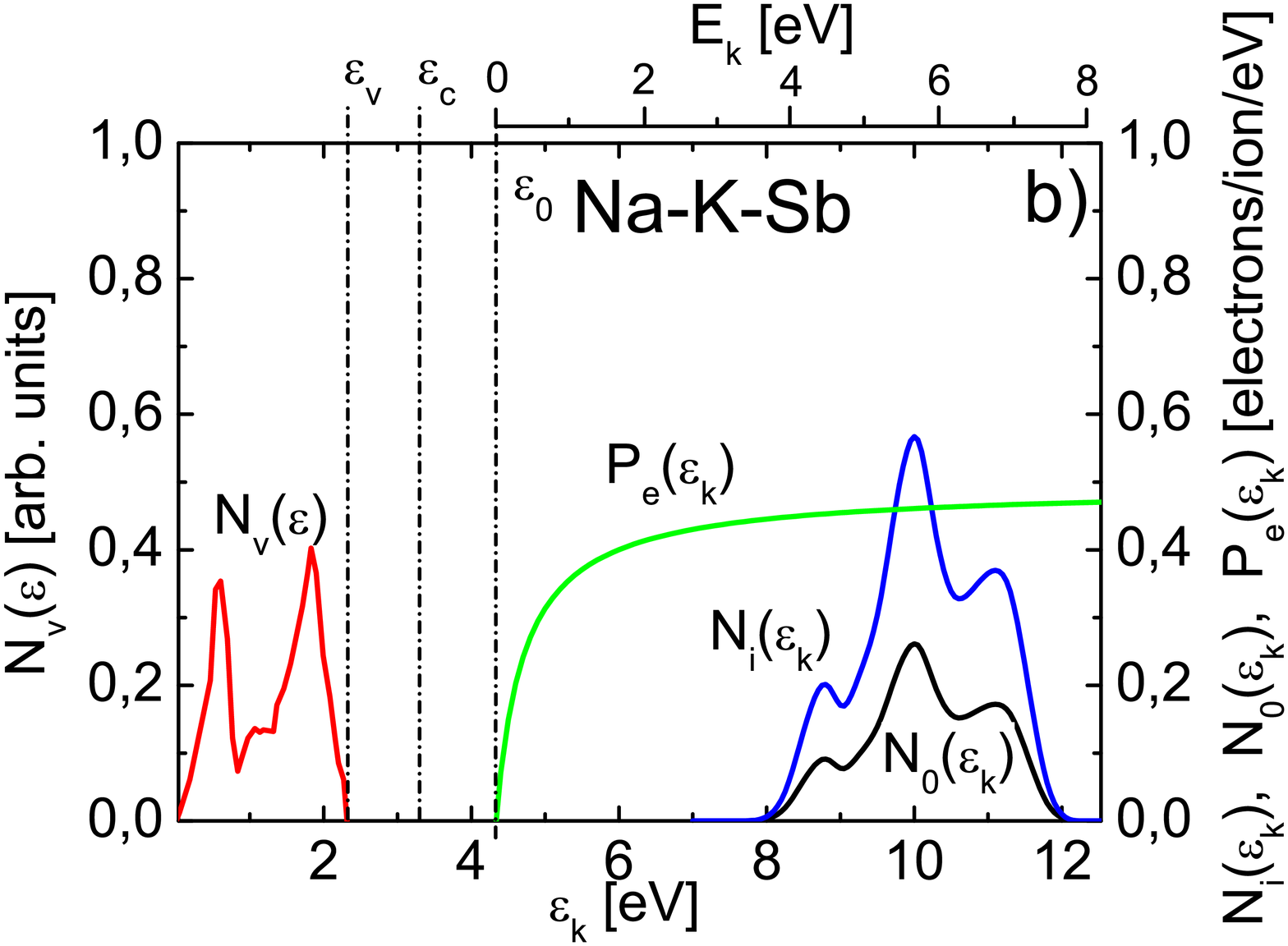}
} \hspace{1cm}
\subfigure 
{
    \label{fig:DOS:NaKSb}
    \includegraphics[width=7cm]{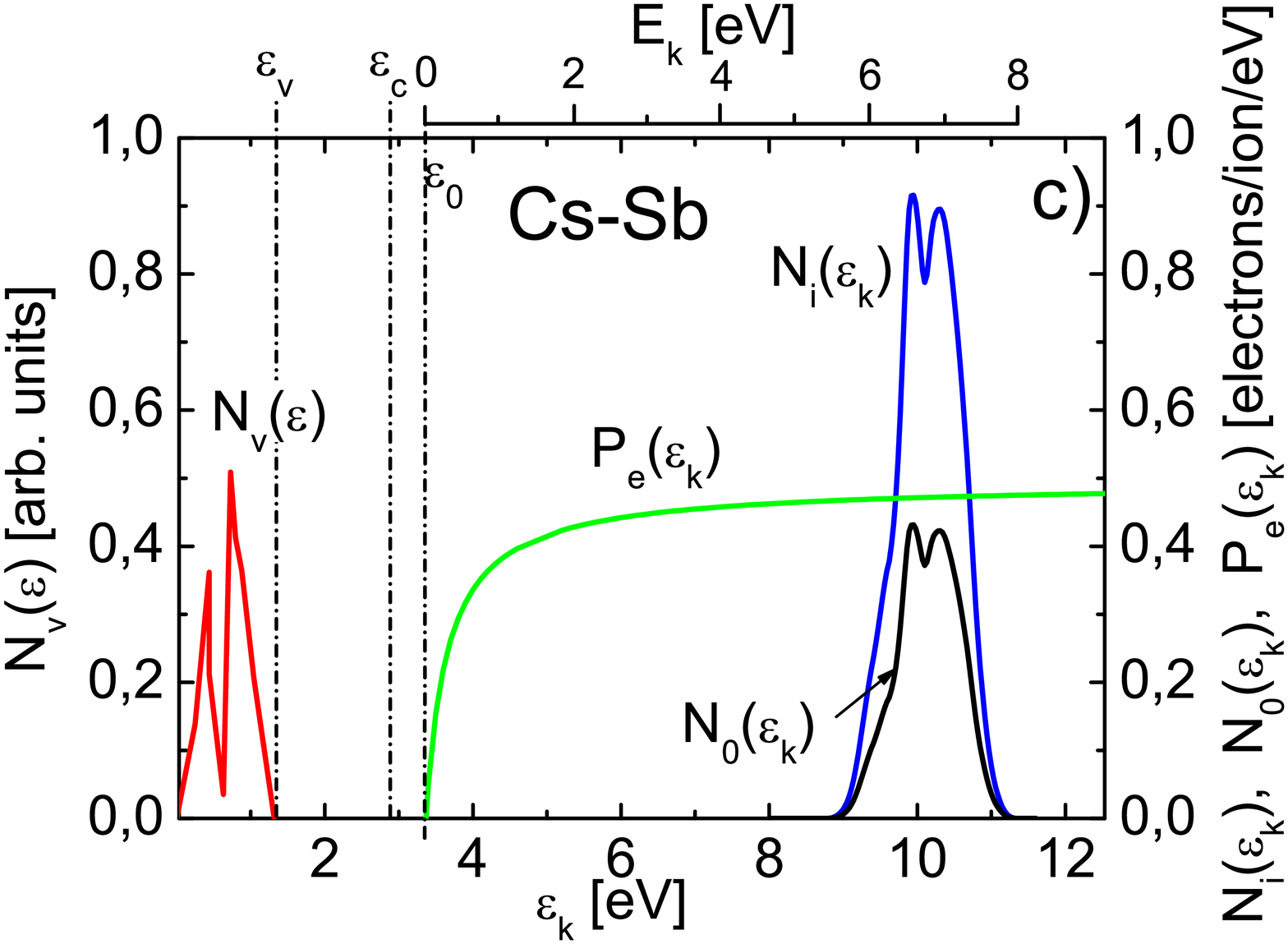}
}
\caption{Plots of density of states $N_v(\varepsilon)$ in the valance band of the semiconductor, kinetic energy distributions $N_i(\varepsilon_k)$ for Auger electrons inside the semiconductor, probability to surmount the vacuum barrier $P_e(\varepsilon_k)$ for Auger electrons as a function of their energies and energy distribution of Auger electrons that escape from the semiconductor $N_0(\varepsilon_k)$. The upper scale indicates kinetic energies of Auger electrons outside the semiconductor $E_k=\varepsilon_{k}-\varepsilon_{0}$. The plots are shown for: a) K-Cs-Sb, b) Na-K-Sb and c) Cs-Sb semiconductor materials.}
\label{fig:DOS:Ne}
\end{figure}

\begin{table}
\begin{ruledtabular}
    \begin{tabular} {lcclcll}
    PC type & $a_{nn}$ & $s_{m}$ & $\kappa$ & $\varepsilon_{v}$ & $\varepsilon_{c}-\varepsilon_{v}$ & $\varepsilon_{0}-\varepsilon_{v}$\\
    & (${\AA}$) & (${\AA}$) & & (eV) & (eV) & (eV)\\ \hline
    K-Cs-Sb & 3.73\footnotemark[1] & 3.76 & 9\footnotemark[2] & 1.27\footnotemark[1] & 1\footnotemark[3] & 1.1\footnotemark[3] \\
    Na-K-Sb & 3.35\footnotemark[4] & 3.57 & 4.66\footnotemark[5] & 2.34\footnotemark[4] & 1\footnotemark[3] & 1\footnotemark[3] \\
    Cs-Sb & 3.95\footnotemark[6] & 3.88 & 3.24\footnotemark[3] & 1.31\footnotemark[7] & 1.6\footnotemark[3] & 0.45\footnotemark[3] \\
    \end{tabular}
    \caption{Parameters used in theoretical calculations for K-Cs-Sb, Na-K-Sb and Cs-Sb photocathodes}
    \label{tab:IISEE:parameters}
\end{ruledtabular}
\footnotetext[1]{from Ref. \cite{ettema:02}}
\footnotetext[2]{from Ref. \cite{motta:05}}
\footnotetext[3]{from Ref. \cite{sommer:80}}
\footnotetext[4]{from Ref. \cite{ettema:00}}
\footnotetext[5]{from Ref. \cite{ebina:73}}
\footnotetext[6]{from Ref. \cite{su-huai:86}}
\footnotetext[7]{from Ref. \cite{tezge:92}}
\end{table}
To calculate the secondary emission coefficient we used the electronic state density function $N_v(\varepsilon)$ in the valence band, which is assumed to be entirely filled. The valence-band state density functions $N_v(\varepsilon)$ for K-Cs-Sb, Na-K-Sb and Cs-Sb bi-alkali compounds respectively, were calculated in \cite{ettema:02, ettema:00, tezge:92} and are schematically depicted in \ref{fig:DOS:Ne}. The energy distribution function $N_i(\varepsilon_k)$ of Auger excited electrons inside the PC is proportional to the product of $N_c(\varepsilon)$, the state density function in empty conduction band, times an Auger transform  $T(\varepsilon)$, which represents the probability to have two electrons in the valence band that can be involved in the process. The latter is thus the integral over the product of state densities $N_v(\varepsilon^{'}) \cdot N_v(\varepsilon^{''})$ in the regions $d\varepsilon^{'}$ and $d\varepsilon^{''}$ at all pairs of energies $\varepsilon^{'}$ and $\varepsilon^{''}$ which are both located at a distance $\Delta$ from $\varepsilon$ in the valence band (\ref{fig:auger:diagram}). With a substitution: $\varepsilon^{'}=\varepsilon-\Delta$, $\varepsilon^{''}=\varepsilon+\Delta$ the Auger transform is then given by:
\begin{equation}
\label{eq:auger:transformation}
      T(\varepsilon)=\int \limits_0^\infty N_v(\varepsilon-\Delta) \cdot N_v(\varepsilon+\Delta) \cdot d \Delta
\end{equation}
Though the integration limit goes to infinity, the integration actually stops when either $\varepsilon-\Delta$ or $\varepsilon+\Delta$ is out of the valence band's boundaries. The state density function $N_c(\varepsilon)$ in the conduction band is assumed to be proportional to the free electron state density function, $(\varepsilon_k-\varepsilon_c)^{1/2}$, for $\varepsilon_k>\varepsilon_c$; it is zero for $\varepsilon_k\leq\varepsilon_c$. Thus, the expression for the energy distribution   of Auger excited electrons inside the PC may be written as:
\begin{equation}
\label{eq:auger:Ni}
      N_i(\varepsilon_k)=K \cdot N_c(\varepsilon_k) \cdot T(\frac{\varepsilon_k+\varepsilon_0-E_i^{'}}{2}),
\end{equation}
where $K$ is a proportionality constant and $\varepsilon=\frac{\varepsilon_k+\varepsilon_0-E_i^{'}}{2}$ is taken from equation \ref{eq:auger:conservation}. K can be evaluated from the normalization of the distribution $N_i(\varepsilon_k)$ to an integral of one electron per Auger neutralized ion:
\begin{equation}
\label{eq:auger:Ni:norm}
      \int \limits_{\varepsilon_c}^\infty N_i(\varepsilon_k)=1
\end{equation}
The shape of $N_i(\varepsilon_k)$ for K-Cs-Sb, Na-K-Sb and Cs-Sb PCs is shown \ref{fig:DOS:Ne}.

We proceed with a calculation of energy distribution $N_0(\varepsilon_k)$ of electrons which leave the PC surface. This calculation requires knowledge about the probability for an excited electron inside the PC to surmount the surface barrier and about the anisotropy of the electron angular distribution inside the PC. The escape probability as a function of kinetic energy is defined as follows \cite{hagstrum:61}:
\begin{eqnarray}
\label{eq:auger:Ni:norm}
      P_e(\varepsilon_k)=\frac{1}{2} \cdot \frac{1-(\varepsilon_0/\varepsilon_k)^{1/2}}{1-\alpha \cdot (\varepsilon_0/\varepsilon_k)^{1/2}}, \quad \varepsilon_k>\varepsilon_0 \nonumber \\ =0, \quad \varepsilon_k<\varepsilon_0
\end{eqnarray}
where $\varepsilon_{0}$ is the height of the surface barrier and the coefficient $\alpha$ reflects the anisotropy of the angular distribution for exited electrons. Hagstrum \cite{hagstrum:61} has determined $\alpha$ to be 0.956 by fitting the theoretical model and experimental data for the case of helium ions neutralized at a Ge surface. The same anisotropy parameters were used by Hagstrum for Ne and Ar ions interacting with either Ge or Si surfaces, thus we use the same value in the following calculations. The escape probability $P_e(\varepsilon_k)$ is plotted in \ref{fig:DOS:Ne} for all PC types used in the calculations.

The energy distribution of electrons which escape the PC bulk $N_0(\varepsilon_k)$ is equal to the product of the energy distribution of Auger excited electrons inside the PC $N_i(\varepsilon_k)$ and the probability $P_e(\varepsilon_k)$ to surmount the surface barrier of hight $\varepsilon_{0}$:
\begin{equation}
\label{eq:auger:N0}
      N_0(\varepsilon_k)=N_i(\varepsilon_k) \cdot P_e(\varepsilon_k).
\end{equation}
The shape of $N_0(\varepsilon_k)$ for K-Cs-Sb, Na-K-Sb and Cs-Sb PCs is shown in \ref{fig:DOS:Ne}.

Finally, the secondary emission probability is expressed as an integral of $N_0(\varepsilon_k)$ over kinetic energies:
\begin{equation}
\label{eq:auger:gamma}
      \gamma_+=\int \limits_{\varepsilon_0}^\infty N_0(\varepsilon_k) \cdot d\varepsilon_k=\int \limits_{0}^\infty N_0(E_k) \cdot dE_k,
\end{equation}
where $E_k=\varepsilon_k-\varepsilon_0$ is the kinetic energy of electrons as they leave the PC.

Thus, we are able to calculate $N_0(\varepsilon_k)$ and $\gamma_+$ as a function of the effective ionization potential   of the incident ion, the electronic state density function $N_v(\varepsilon)$ in the valence band and the energy-band parameters $\varepsilon_v$, $\varepsilon_c$ and $\varepsilon_0$ of the semiconductor (see table \ref{tab:IISEE:parameters}).

Clearly, the secondary emission coefficient   calculated above is referred to vacuum environment. For PC operation in gas media, the measured secondary emission coefficient will be different, due to the scattering of the Auger electrons by gas molecules back to the PC \cite{breskin:02, moermann:thes}. We denote the ion-induced secondary emission coefficient in a gas media as $\gamma_+^{eff}$; it is given by:
\begin{equation}
\label{eq:auger:gamma:eff}
      \gamma_+^{eff}=\gamma_+ \cdot \varepsilon_{extr}
\end{equation}
where $\varepsilon_{extr}$ is the fraction of secondary Auger electrons which were not backscattered. Thompson equation \cite{theobald:53, sahi:76} estimates the fraction $\varepsilon_{extr}$ as
\begin{equation}
\label{eq:extr:thompson}
      \varepsilon_{extr}=\frac{4 \cdot v_d}{\overline{v_{ae}}+4 \cdot v_d}
\end{equation}
where $v_d$ is the electron drift velocity in the gas ($v_d=2.34\cdot10^4$ m/s \cite{becker} in Ar/CH$_{4}$ (95/5) gas mixture for 273K and 700 Torr at an electric field of 0.5 kV/cm) and $\overline{v_{ae}}$ is the mean velocity of secondary Auger electrons emitted from the PC. This expression provides a fair estimate if the average kinetic energy of the emitted electrons is higher than the average equilibrium kinetic energy of electrons in the gas \cite{theobald:53, burch:77}. In the case of Ar/CH$_4$ (95/5) at 700 Torr and drift field of 0.5kV/cm, the average equilibrium kinetic energy for electrons is about 2 eV \cite{sebastian:05} which is smaller than the average kinetic energy of Auger electrons for the PCs investigated (\ref{fig:DOS:Ne}); therefore equation \ref{eq:extr:thompson} is valid and should provide a good estimate for $\varepsilon_{extr}$. The average velocity of the emitted Auger electrons was evaluated from the $N_0(\varepsilon_k)$ energy distributions: first the average kinetic energy $\overline{E_k}$ was calculated by the equation\[\overline{E_k}=\int \limits_0^\infty N_0(E_k)\cdot E_k \cdot dE_k / \int \limits_0^\infty N_0(E_k)\cdot dE_k\;; \]
the mean velocity is then given by:
\begin{equation}
\label{eq:auger:kinetic:average}
      \overline{v_{ae}}=\sqrt{\frac{2 \cdot \overline{E_k}}{m_e}}
\end{equation}
where $m_e$ is the electron mass. The calculated average velocities for Auger electrons are: $1,51\cdot10^6$ m/s for K-Cs-Sb, $1,43\cdot10^6$ for Na-K-Sb and $1,54\cdot10^6$ m/s for Cs-Sb.
The calculated values of $\varepsilon_{extr}$ and $\gamma_+^{eff}$ for K-Cs-Sb, Na-K-Sb and Cs-Sb PCs are listed in Table \ref{tab:IISEE:calculations}.

\begin{table}
\begin{ruledtabular}
    \begin{tabular} {ccccc}
    PC type & $E_i^{'} [eV]$ & $\gamma_+$ & $\varepsilon_{extr}$ & $\gamma_+^{eff}$ \\ \hline
    K-Cs-Sb & 11.83 & 0.471 & 0.0583 & 0.0274 \\
    Na-K-Sb & 11.95 & 0.491 & 0.0614 & 0.030 \\
    Cs-Sb & 12.11 & 0.472 & 0.0572 & 0.0270 \\
    \end{tabular}
    \caption{Calculated $E_i^{'}$, $\gamma_+$, $\varepsilon_{extr}$ and $\gamma_+^{eff}$ values for K-Cs-Sb, Na-K-Sb and Cs-Sb photocathodes.}
    \label{tab:IISEE:calculations}
\end{ruledtabular}
\end{table}

The calculated values $\varepsilon_{extr}$ (Table \ref{tab:IISEE:calculations}) indicates that about 94\% of Auger electrons are scattered back to the PC in Ar/CH$_{4}$ (95/5) at 700 mbar and 0.5kV/cm.

Equation \ref{eq:extr:thompson} is not applicable for the photoelectron case as their average kinetic energies are approximately between 0.7 eV and 1.2 eV \cite{nathan:70}, a range which is lower than the average equilibrium kinetic energy ($\sim$2 eV) of electrons in the gas at the conditions mentioned above. Data on the backscattering probability of photoelectrons in various gases may be found in \cite{breskin:02, escada:07, coelho:07}.

\subsection{\label{sec:Gmeas:theory}Estimation of IISEE effects in visible-sensitive GPMs}

\begin{figure} [t]
  \makeatletter
    \renewcommand{\p@figure}{Fig.\space}
  \makeatother
    \includegraphics[width=7cm]{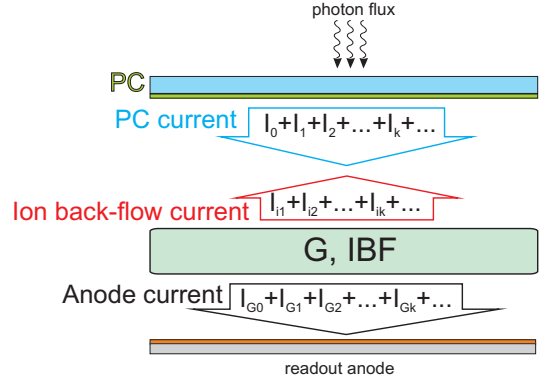}
    \caption{Operation of a visible sensitive GPM: a gaseous multiplier characterized by a multiplication factor G and a fraction of avalanche ions reaching the PC ($IBF$) is coupled to a visible sensitive PC. The PC is constantly illuminated by a light source inducing a primary photoelectron current $I_{0}$; as the photoelectrons are multiplied in the multiplier, a current $I_{G0}$ is read at the readout anode while the primary current of avalanche ions $I_{i1}$ is read at the PC. }
    \label{fig:ifeedback:development}
\end{figure}

The contribution of IISEE to the anode current recorded in a GPM can be estimated from the electron multiplication and the secondary electron emission. We assume that the PC is illuminated with a constant light flux. The average multiplication coefficient (gain) is denoted by $G$ and the fraction of avalanche-induced ions flowing back to the PC is denoted by $IBF$; both are characteristics of the multiplier's structure and operation voltages. The light induces a photo-current $I_{0}$ at the PC, which, assuming full photoelectron collection to the multiplier (usually the case in GEM multipliers), yields after multiplication a current equal to $I_{G0}=I_{0} \cdot G$  at the anode (see \ref{fig:ifeedback:development}). The current of back-flowing ions reaching the PC, which are produced by these avalanches, is $I_{i1}=I_{0} \cdot G \cdot IBF$. These ions impinging on the PC surface have a probability $\gamma_+$ to produce secondary Auger electrons; a fraction $\varepsilon_{extr}$ of them will be emitted from the PC and initiate, after multiplication, a secondary-electron anode current $I_{G1}=I_{0} \cdot G^2 \cdot IBF \cdot \gamma_+ \cdot \varepsilon_{extr}$. This, in turn, induces a second generation of back-drifting ions, further Auger-electrons production at the PC, etc.; the process results in a decreasing geometric series of currents ($I_{G1}=I_{0} \cdot G^2 \cdot IBF^2 \cdot \gamma_+ \cdot \varepsilon_{extr}$ , $I_{G3}=I_{0} \cdot G^3 \cdot (IBF \cdot \gamma_+ \cdot \varepsilon_{extr})^2$, and so on (see \ref{fig:ifeedback:development}). A condition: $IBF \cdot \gamma_+ \cdot \varepsilon_{extr} \cdot G<1$ is required to avoid the series divergence. The k-th contribution of the IISEE can be formulated as $I_{Gk}=I_{0} \cdot G^{k+1} \cdot (IBF \cdot \gamma_+ \cdot \varepsilon_{extr})^k$. The total anode current equals to the sum of all contributions, given by:

\begin{eqnarray}
\label{eq:ifeedback:current1}
      I_A=I_{G0}+I_{G1}+I_{G2}...=\sum \limits_{k=1}^{\infty} I_{Gk}= \nonumber\\ =I_0 \cdot G \cdot \sum \limits_{k=1}^{\infty} (G \cdot IBF \cdot \gamma_+ \cdot \varepsilon_{extr})^k
\end{eqnarray}

\noindent which may be also written as:

\begin{eqnarray}
\label{eq:ifeedback:current2}
      I_A=I_0 \cdot G \cdot \sum \limits_{k=1}^{\infty} (G \cdot IBF \cdot \gamma_+ \cdot \varepsilon_{extr})^k= \nonumber\\ =I_0 \cdot \frac{G}{1-G \cdot IBF \cdot \gamma_+ \cdot \varepsilon_{extr}},
\end{eqnarray}

\noindent or, in terms of gain,

\begin{equation}
\label{eq:ifeedback:gain}
      G_{meas}=\frac{G}{1-G \cdot IBF \cdot \gamma_+ \cdot \varepsilon_{extr}},
\end{equation}

\noindent where the measured gain, $G_{meas}$ is the ratio of the measured anode current $I_A$ to the primary photocurrent $I_0$ (\ref{fig:IISEE:setup} and \ref{fig:ifeedback:development}), and $I_0$ is the multiplier's gain in the absence of ion feedback. To remind, IBF, the fraction of avalanche ions reaching the PC, is measured by the ratio of the PC current $I_{PC}$ under avalanche multiplication to the anode current $I_A$ (\ref{fig:IISEE:setup} and \ref{fig:ifeedback:development}).

\subsection{\label{sec:IISEE:exp}Experimental determination of the IISEE effects}

\begin{figure}
    \includegraphics[width=7cm]{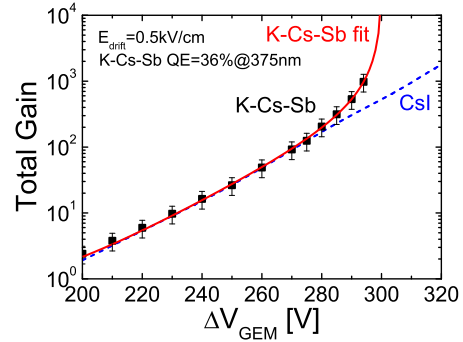}
    \caption{Gain-voltage characteristics measured in the GPM of \protect \ref{fig:IISEE:setup} (see conditions in the figure, QE refers to vacuum) with CsI (dashed line) and K-Cs-Sb (open circles) photocathodes. The divergence from exponential with K-Cs-Sb is due to ion-induced secondary electron emission. The solid line is a fit to the experimental data points using equation \ref{eq:ifeedback:gain}.}
    \label{fig:2G:gain:KCsSb}
\end{figure}

\begin{figure}
  \makeatletter
    \renewcommand{\p@figure}{Fig.\space}
  \makeatother
    \includegraphics[width=7cm]{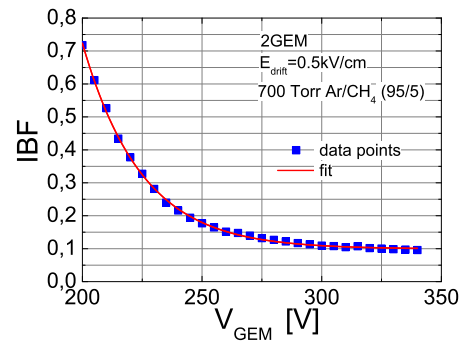}
    \caption{Ion back-flow fraction ($IBF$) as a function of GEM voltage for the double-GEM multiplier shown in \protect \ref{fig:IISEE:setup}, measured in 700 Torr of Ar/CH$_4$ (95/5). The solid line is a fit to the data points.}
    \label{fig:2G:IBF}
\end{figure}

The effective probability of IISEE defined above, $\gamma_+^{eff}$, can be extracted from the measured gain with IISEE, $G_{meas}$, if $G$ (measured gain without IISEE) and IBF are known. Normally, G is an exponential curve, and the IISEE will be manifested as a deviation from this exponent (equation \ref{eq:ifeedback:gain}). As an example, \ref{fig:2G:gain:KCsSb} shows a gain curve obtained with a double-GEM coupled to a semi-transparent K-Cs-Sb PC, as function of $\Delta V_{GEM}$ (\ref{fig:IISEE:setup}). Up to $\Delta V_{GEM}=280$ V the gain increases exponentially and above 300V it diverges. At this point the quantity $G \cdot IBF \cdot \gamma_+ \cdot \varepsilon_{extr}$ in equation equation \ref{eq:ifeedback:gain} approaches unity, leading to detector's break-down. For comparison, a second gain curve (dashed line) is plotted in \ref{fig:2G:gain:KCsSb}, obtained with the same detector under the same operation conditions, but coupled to a semi-transparent CsI PC; the parameter G as a function of $\Delta V_{GEM}$ can be derived from this curve. As IISEE in CsI is negligible, due to a very wide band-gap of about 6 eV, the gain curve grows exponentially even at the highest operation potentials. The IBF as a function of GEM voltage was measured in the same detector (geometry, gas and voltages), with a CsI PC, and was plotted in \ref{fig:2G:IBF} as function of $\Delta V_{GEM}$. The data points in \ref{fig:2G:IBF} were fitted with an exponential function, which seems to appropriately describe the dependence of IBF on the GEM voltage. With the known dependence of $IBF$ and $G$ (properties of the multiplier that are independent of the PC) on the GEM voltage $\Delta V_{GEM}$, the parameter   could be derived (using equation \ref{eq:ifeedback:gain}) from the gain-voltage curve of the same multiplier coupled to a visible-sensitive PC. This procedure, however, has a large uncertainty; namely, an inaccuracy in adjustment of the total gain, which was very sensitive to small voltage deviations, resulting in $\sim$30\% error in $\gamma_+^{eff}$.

\begin{figure} [t]
    \includegraphics[width=7cm]{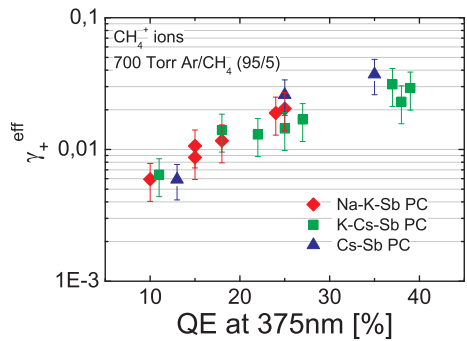}
    \caption{Measured effective probability of the ion-induced secondary electron emission coefficient for K-Cs-Sb, Na-K-Sb and Cs-Sb PCs as a function of their quantum efficiency values (QEs).}
    \label{fig:gamma:QE}
\end{figure}

The gain-voltage curves were measured for several samples of various visible-sensitive PCs: eight K-Cs-Sb samples, six Na-K-Sb samples and three Cs-Sb PCs. As the emission properties of a given PC type varied from sample to sample, we could establish a significant data base of $\gamma_+^{eff}$ - values as function of the QE. The correlation is shown in \ref{fig:gamma:QE}. As one can see, $\gamma_+^{eff}$ increases with QE, reaching a value of about 0.03 electrons/ion for the most efficient PCs; its value is independent of the PC type.

\section{\label{sec:discussion}Discussion}

Experimental and theoretical approaches were undertaken to estimate the probability of ion-induced secondary electron emission in a gas medium; the research work is within our ongoing efforts to develop gas-avalanche photomultipliers sensitive to single photons in the visible spectral range.

A simple theoretical model was adopted for calculating the ion-induced secondary emission coefficient $\gamma_+^{eff}$ from bi- and mono-alkali photocathodes. We assumed the Auger neutralization process as the main mechanism for the secondary electron emission and used basic properties of semiconductors to evaluate the emission probability. The input parameters for the calculation were the effective ionization potential $E_i^{'}$ of the incident ion, the density of states in the valence band and the energy-band parameters $\varepsilon_v$, $\varepsilon_c$ and $\varepsilon_0$  of the semiconductor (the PC). $\gamma_+$ was calculated  to be 0.47, 0.49 and 0.47 for K-Cs-Sb, Na-K-Sb and Cs-Sb PCs, correspondingly. The calculated values of $\gamma_+$ are quite similar for the three PCs, which can be explained by the similarity of their energy-band parameters. The bi- and mono-alkali PC materials used here have a very narrow valence band (1-2 eV) as compared to the effective ionization energy of methane ions ($\sim$12 eV). This leads to narrow (2-4 eV) energy distributions of the Auger electrons (\ref{fig:DOS:Ne}), peaking at a rather high energy of about 6 eV (top scale of \ref{fig:DOS:Ne}) the energies of Auger electrons are in a range from 4 eV to 8 eV.

The effective secondary emission coefficient $\gamma_+^{eff}$, namely the one corrected for the backscattering, in Ar/CH$_4$ (95/5) at 700 mbar and field 0.5 kV/cm was calculated from the theoretical model to be 0.027 for Cs-Sb and K-Cs-Sb PCs and 0.03 for Na-K-Sb PC.

A simple method for experimental extraction of $\gamma_+^{eff}$ from measured voltage-gain curves was introduced. It is based on recording and comparing avalanche currents on the PC and anode of a double-GEM multiplier coupled to CsI and visible-sensitive PCs, under the same operation conditions. CsI, with no ion feedback, provided the multiplication factor and the ion back-flow fraction, while the visible-sensitive PCs provided the effective secondary emission coefficient into gas, $\gamma_+^{eff}$, derived from a fit of the gain-voltage curve through equation \ref{eq:ifeedback:gain}. $\gamma_+^{eff}$ in Ar/CH$_4$ (95/5) at 700 Torr were between 0.02 and 0.03 Auger electrons per incident ion for Na-K-Sb, K-Cs-Sb and Cs-Sb photocathodes; it is in good agreement with our theoretical estimations, despite the fact that the experimental PC surfaces were most probably not perfect ones from the point of view of stoichiometry \cite{shefer:thes}, defects and traps. For all three PCs investigated, $\gamma_+^{eff}$ increased with the PC's QE, regardless of the PC type. It is therefore reasonable to believe that the theoretical calculations yielded a reasonable estimate of $\gamma_+^{eff}$ only for the highest QE-values.

Based on our $\gamma_+^{eff}$ - values, we can analyze the requirements on any type of gaseous multiplier used in combination with a visible-sensitive PC (in the present gas); other gases would require the calculation of the electron backscattering coefficients. In particular the ion back-flow fraction and gain permitting stable continuous-mode operation of visible-sensitive gaseous photomultipliers can be estimated, through equation \ref{eq:ifeedback:gain}. To avoid avalanche divergence into a spark it should fulfill the condition:  $G \cdot IBF \cdot \gamma_+ \cdot \varepsilon_{extr}<1$.  Thus, a gain of 10$^5$ (e.g. required for high single-photon sensitivity in GPMs) and  $\gamma_+^{eff}$=0.03 imply $IBF<3,3\cdot 10^{-4}$.

\begin{figure}
    \includegraphics[width=7cm]{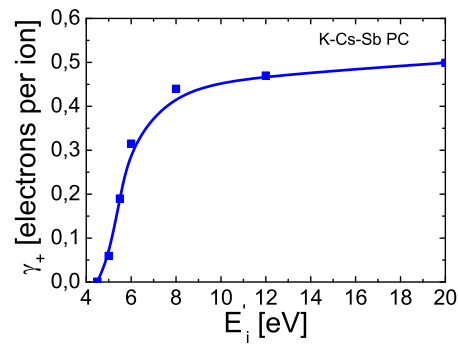}
    \caption{Calculated dependence of the secondary emission coefficient in K-Cs-Sb on the effective ionization energy.}
    \label{fig:gamma:Ei}
\end{figure}

Further reduction of the IISEE coefficient, $\gamma_+$, for these bi- and mono-alkali photocathodes, could only be envisaged by using other gases, with lower effective ionization energy than Ar/CH$_4$ (95/5). As an illustration, the dependence of $\gamma_+$ for K-Cs-Sb on the effective ionization energy of the ion was calculated using the above model, and is presented in \ref{fig:gamma:Ei}. A two-fold reduction of $\gamma_+$, as compared to the value of 0.47 calculated for methane ions, would require a gas with effective ionization energy of about 6 eV. Some low ionization potential photosensitive vapors like triethylamine (TEA, ionization potential 7.5 eV) \cite{peskov:80, charpak:79}, tetrakis(dimethylamine)ethylene (TMAE, ionization potential 5.36 eV) \cite{anderson:80} and ethyl ferrocene (EF, ionization potential 6.1 eV) \cite{charpak:89} employed in some gaseous detectors could be admixed to the GPM's gas filling in order to decrease the IISEE probability. For example, EF yielded stable operation of a detector with CsI photocathode \cite{charpak:93}. It was shown \cite{biteman:01} that in the GPM filled with He/CH$_4$ comprised of Cs-Sb PC and a capillary plate, an addition of EF vapor to the gas mixture slightly improved the PC's QE and the maximal achievable gain of the device. Other gases with low ionization potentials ($\sim$10 eV) employed in detectors are long-chain hydrocarbons, e.g. iso-butane. The ions of such gases have a high probability for dissociation, creating free radicals; they are known to induce aging in gas avalanche detectors. The photocathodes may suffer from enhanced chemical aging when operated in such an environment.

\begin{figure}[h]%
\subfiguretopcaptrue
\subfigure 
{
    \label{fig:Xsections:Ar}
    \includegraphics[width=7cm]{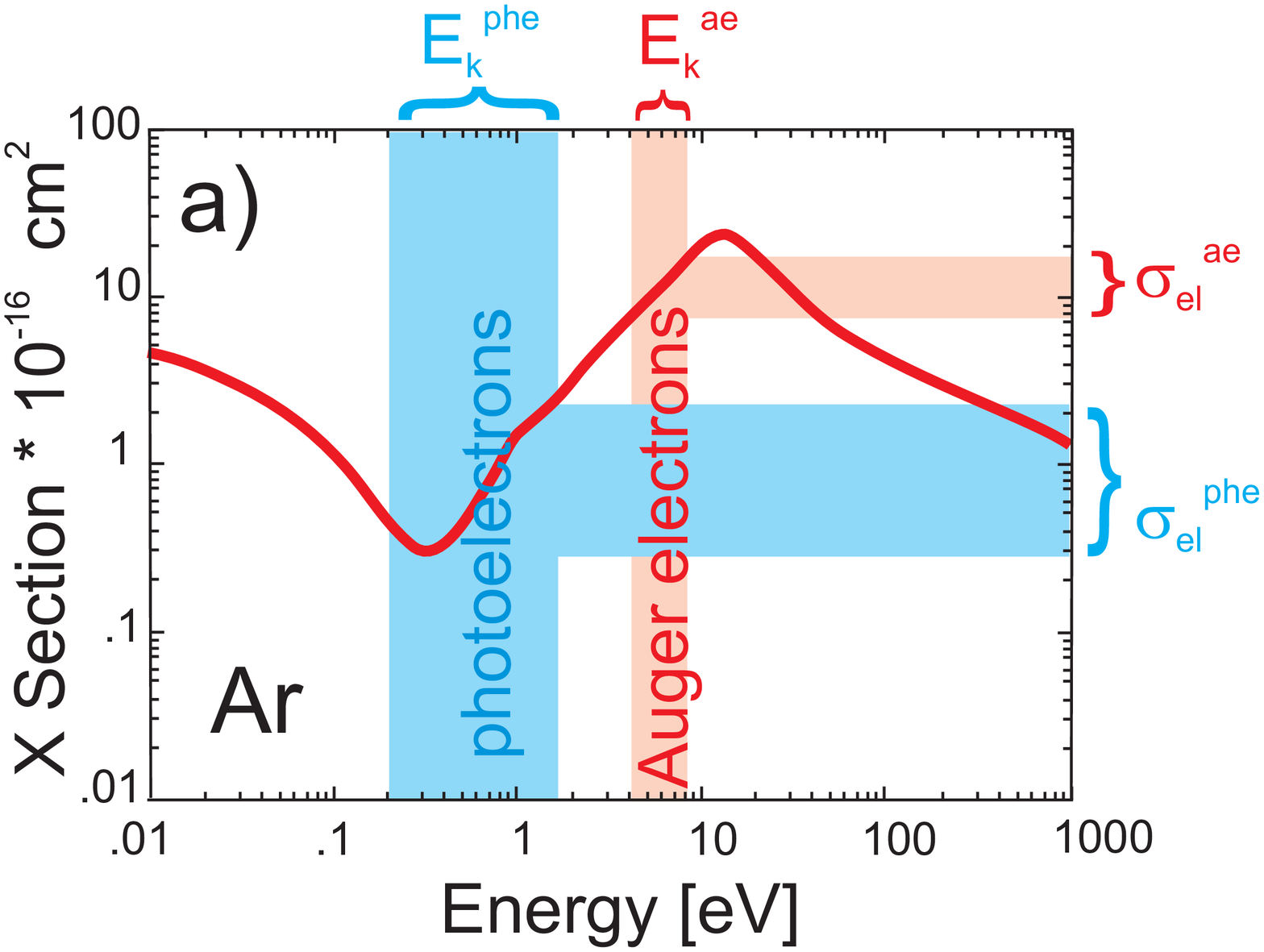}
} \hspace{0.2cm}
\subfigure 
{
    \label{fig:Xsections:CH$_4$}
    \includegraphics[width=7cm]{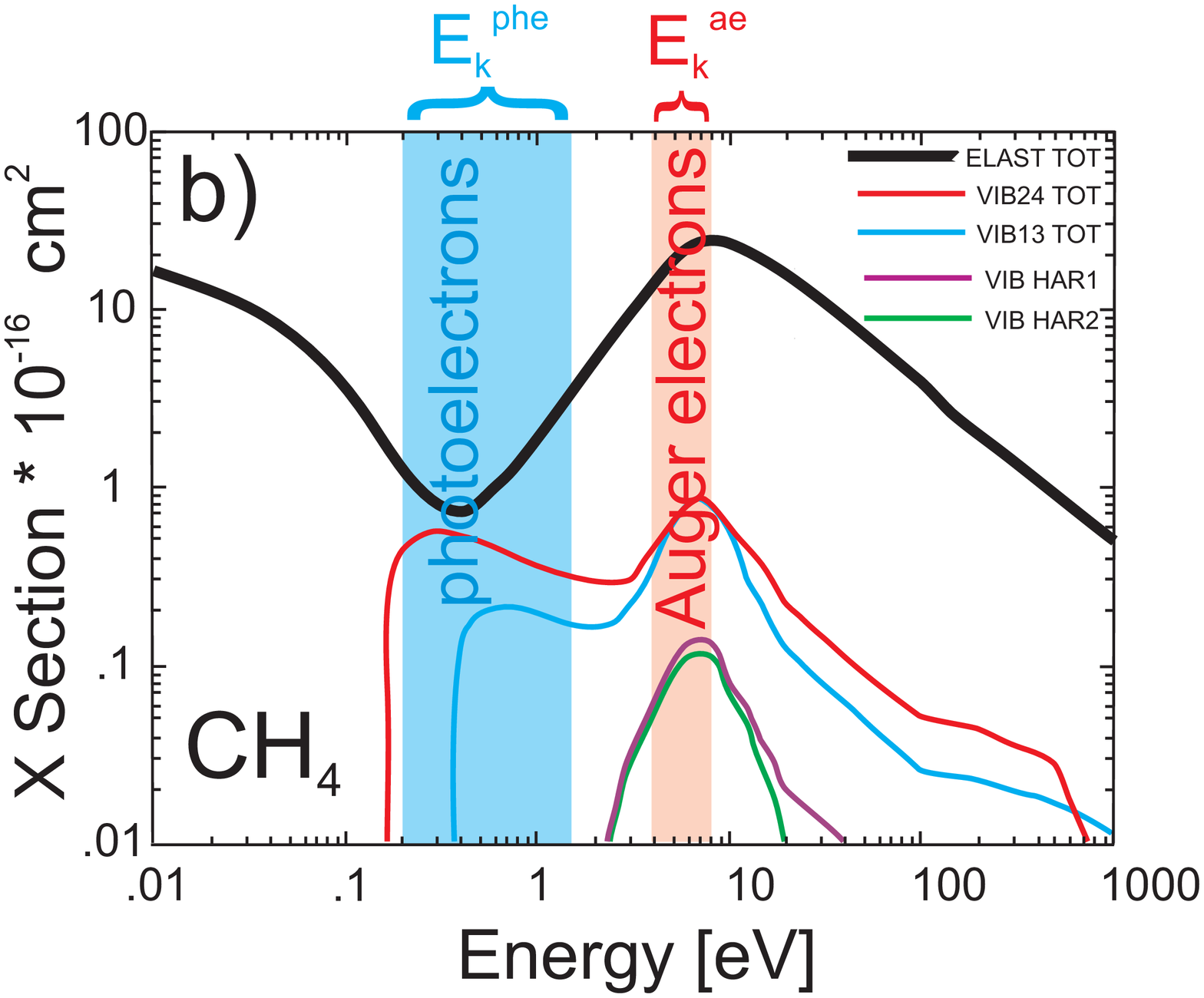}
} \caption{a) Elastic scattering cross-sections for argon. The photoelectron's energies are in the Ramsauer minimum of the elastic cross-section curves, while the Auger electrons energies correspond to rather high cross-sections. b) Elastic and inelastic cross-sections for methane. The photoelectron $E_k^{phe}$ and Auger electron $E_k^{ae}$ energy spreads are shown. The ratio of elastic/inelastic cross-sections is much smaller for photoelectrons than that for Auger electrons. The plots were taken from an open cross-section database at \protect \cite{magboltz}.
}
\label{fig:Xsections}
\end{figure}

A further reduction of the effective secondary emission coefficient $\gamma_+^{eff}=\gamma_+ \cdot \varepsilon_{extr}$, (the fraction of Auger electrons that surmounted the backscattering with gas molecules) may be obtained by increasing the Auger electrons' backscattering, while keeping it low for photoelectrons. Due to the significant difference in the energy spectrum of Auger- and photo-electrons, it might be possible to choose a gas that complies to this requirement, as indicated in \ref{fig:Xsections}.

\begin{figure}
    \includegraphics[width=7cm]{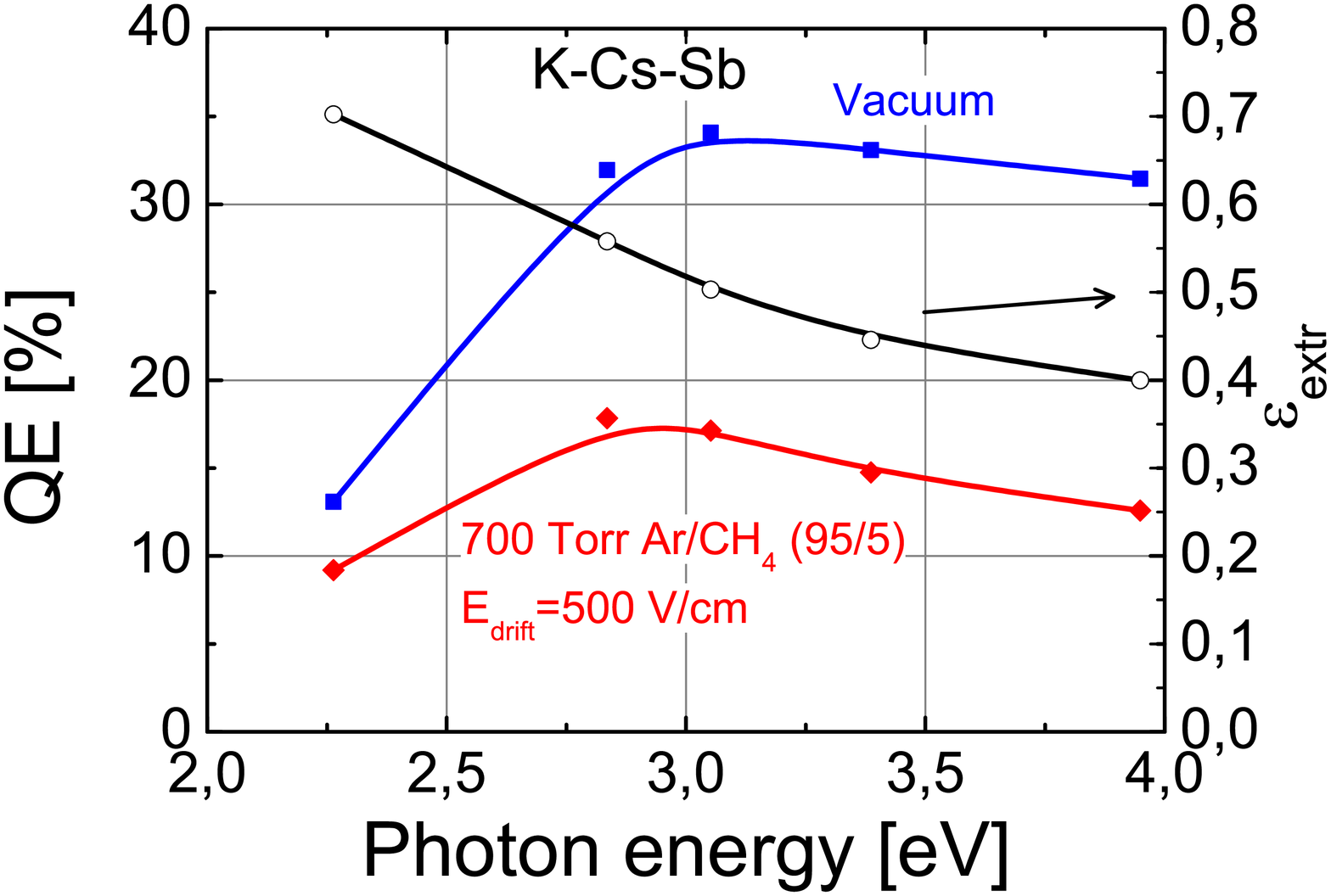}
    \caption{A typical plot of QE as a function of photon energy for a K-Cs-Sb PC, measured in vacuum and in 700 Torr of Ar/CH$_{4}$ (95/5) with a drift field of 500V/cm. The fraction $\varepsilon_{extr}$ of photoelectrons surmounted backscattering in the gas as a function of wavelength is also shown (circles).}
    \label{fig:KCsSb:extr:eV}
\end{figure}

In the corresponding kinetic energy range between 4 eV and 8 eV, the fraction of Auger electrons which scattered back to the PC following collisions with gas molecules was estimated with the model to be rather high, about $\sim$94\% in Ar/CH$_4$ (95/5) at 700 Torr, under a drift field of 0.5kV/cm; for comparison, the photoelectrons (induced by UV-to-visible light photons in the range between 2.2 eV and 4 eV) backscattering in the same conditions was measured to be in a range between 60\% and 20\% for K-Cs-Sb PC (\ref{fig:KCsSb:extr:eV}). The large difference in backscattering for Auger electrons and photoelectrons is apparently due to the difference in their initial kinetic energies. The kinetic energy distributions for photoelectrons originated from K-Cs-Sb by visible-range photons (2.1 eV to 3.1 eV) are peaked between 0.7 eV and 1.2 eV,  extending from 0.2 to 1.7 eV  \cite{nathan:70}; the distributions for Na-K-Sb and Cs-Sb are assumed to be essentially similar. The Auger electrons induced by ions are more energetic according to our calculations and their energies extend between 4 and 8 eV (see \ref{fig:DOS:Ne}). On the other hand, the electron scattering cross-sections for argon and methane are functions of the electron's kinetic energy; they are depicted in \ref{fig:Xsections:Ar} and \ref{fig:Xsections:CH$_4$}. For noble gases, higher backscattering is related to higher elastic scattering cross-section (and to smaller energy loss in elastic collisions, when going from light to heavy gas); for more complex molecules, the backscattering behaves as the ratio of elastic-to-inelastic scattering, vibrational excitation collisions playing an important role in cooling down the energy of the photoelectrons in the gas \cite{dimauro:96, dias:04, escada:07}.

In the energy range between 4 and 8 eV (Auger electrons) either the elastic scattering cross-sections (argon) or the ratios of elastic-to-inelastic scattering cross-sections (methane) are rather large and so is the backscattering; photoelectrons of energies ranging between 0.2 and 1.7 eV fall into the region of Ramsauer minimum of elastic cross-sections for argon; in methane there is also a Ramsauer minimum at about the same energy as in argon, in the vicinity of which the ratio of elastic/inelastic cross-sections is small; the corresponding back-scattering is several times smaller than that for Auger electrons.

Choosing a GPM's operating gas in which backscattering is high for Auger electrons and low for photoelectrons will result in good photoelectron extraction (high effective QE) and reduced probability of ion-induced secondary effects. This point to a choice of gases with pronounced Ramsauer minimum located close to the energy range of photoelectrons induced by visible light photons. Examples are heavy noble gases, namely Xe, Kr, Ar and methane. A gas mixture of Ar and CH4 exhibiting this property, was used in the present work.

\bibliography{publications}

\end{document}